\documentclass[aps,prl,reprint,longbibliography,superscriptaddress]{revtex4-1}

\usepackage{graphicx}
\usepackage{amssymb,amsmath,amsthm,bm}
\usepackage{physics}
\usepackage{lipsum}
\usepackage{color}
\usepackage{float}
\usepackage{amsmath}
\usepackage{xcolor}
\definecolor{myred}{RGB}{216,18,116} 
\usepackage[colorlinks=true,linkcolor=myred,citecolor=myred, pdfauthor={ },pdftitle={ },pdfsubject={ },pdfkeywords={ }]{hyperref}
\usepackage[export]{adjustbox}

\begin{document}

\title{Quantum Simulation of Spin-Dependent Electron Transfer in a Synthetic Chiral Lattice with a Trapped Ion}

\author{Yi Li}
\thanks{These authors contribute equally to this work.}
\affiliation{Laboratory of Spin Magnetic Resonance, School of Physical Sciences,
Anhui Province Key Laboratory of Scientific Instrument Development and Application,
University of Science and Technology of China, Hefei 230026, China}
\affiliation{National Advanced Talent Cultivation Center for Physics, University of Science and Technology of China, Hefei 230026, China}
\affiliation{Hefei National Laboratory, University of Science and Technology of China, Hefei 230088, China}

\author{Chuyuan Chen}
\thanks{These authors contribute equally to this work.}
\affiliation{Laboratory of Spin Magnetic Resonance, School of Physical Sciences,
Anhui Province Key Laboratory of Scientific Instrument Development and Application,
University of Science and Technology of China, Hefei 230026, China}

\author{Xingyu Zhao}
\affiliation{Laboratory of Spin Magnetic Resonance, School of Physical Sciences,
Anhui Province Key Laboratory of Scientific Instrument Development and Application,
University of Science and Technology of China, Hefei 230026, China}
\affiliation{Hefei National Laboratory, University of Science and Technology of China, Hefei 230088, China}

\author{Zihan Xie}
\affiliation{Laboratory of Spin Magnetic Resonance, School of Physical Sciences,
Anhui Province Key Laboratory of Scientific Instrument Development and Application,
University of Science and Technology of China, Hefei 230026, China}
\affiliation{Hefei National Laboratory, University of Science and Technology of China, Hefei 230088, China}

\author{Min Jiang}
\affiliation{Laboratory of Spin Magnetic Resonance, School of Physical Sciences,
Anhui Province Key Laboratory of Scientific Instrument Development and Application,
University of Science and Technology of China, Hefei 230026, China}

\author{Xinhua Peng}
\affiliation{Laboratory of Spin Magnetic Resonance, School of Physical Sciences,
Anhui Province Key Laboratory of Scientific Instrument Development and Application,
University of Science and Technology of China, Hefei 230026, China}
\affiliation{Hefei National Laboratory, University of Science and Technology of China, Hefei 230088, China}
\affiliation{Hefei National Research Center for Physical Sciences at the Microscale,
University of Science and Technology of China, Hefei 230026, China}

\author{Han Pu}
\affiliation{Department of Physics and Astronomy, and Smalley-Curl Institute, Rice University, Houston, TX 77005, USA}

\author{Lyuzhou Ye}
\affiliation{Hefei National Research Center for Physical Sciences at the Microscale, University of Science and Technology of China, Hefei 230026, China}

\author{Yao Wang}
\affiliation{Hefei National Research Center for Physical Sciences at the Microscale, University of Science and Technology of China, Hefei 230026, China}

\author{Guozhen Zhang}
\email{guozhen@ustc.edu.cn}
\affiliation{Hefei National Laboratory, University of Science and Technology of China, Hefei 230088, China}
\affiliation{Hefei National Research Center for Physical Sciences at the Microscale, University of Science and Technology of China, Hefei 230026, China}

\author{Yiheng Lin}
\email{yiheng@ustc.edu.cn}
\affiliation{Laboratory of Spin Magnetic Resonance, School of Physical Sciences,
Anhui Province Key Laboratory of Scientific Instrument Development and Application,
University of Science and Technology of China, Hefei 230026, China}
\affiliation{Hefei National Laboratory, University of Science and Technology of China, Hefei 230088, China}
\affiliation{Hefei National Research Center for Physical Sciences at the Microscale,
University of Science and Technology of China, Hefei 230026, China}


\begin{abstract}
Electron transfer through chiral structures can exhibit spin asymmetry, 
known as the chiral-induced spin selectivity effect, whose microscopic origin remains an open question. 
While path-interference within the chiral moiety has been proposed as a key mechanism, its experimental validation requires precise and versatile tunability of system parameters. 
Here we implement a programmable quantum simulation of spin-dependent electron transfer in a donor~-~chiral bridge~-~acceptor model using a trapped ion. The bridge is encoded in internal states of the ion with tunable nearest- and next-nearest-neighbor couplings, while donor and acceptor states are coupled via a spectator bosonic motional mode. We observe spin-dependent interference within the bridge, and further reveal spin-dependence in donor-to-acceptor transfer dynamics, controlled by amplitude and phase of the coupling parameter. Our results identify interference among spin-dependent pathways as a microscopic origin of spin-dependent transfer, and open a route toward quantum simulations of complex chiral lattices with multi-level and bosonic degrees of freedom.

\end{abstract}

\maketitle
Chirality is a fundamental structural property that plays a central role across physics, chemistry, and biology systems~\cite{yan2024structural,kelvin1894molecular,blackmond2011origin}. In particular, it underlies a wide range of phenomena in molecular systems, from stereoselective reactions~\cite{Krautwald_stereodivergence_2017, wang_construction_2018, Cheng_recent_2021} to charge and energy transport in complex environments~\cite{brandt2017added,hu2020chiral,zhao2020new,zhong2024matching}. 
For electron transfer 
through chiral structures, experiments have revealed spin-dependent transport, known as the chiral-induced spin selectivity (CISS) effect~\cite{naaman2012chiral, naaman2019chiral, bloom2024chiral}. This effect has been observed on various experimental platforms~\cite{Ray1999Asymmetric,xie2011spin,Hannah2023} and has attracted significant attention for its potential applications in spintronics and quantum devices based on chirality~\cite{naaman2015spintronics, yang2021chiral,aiello2022chirality}. However, the microscopic origin of CISS remains unresolved~\cite{evers2020advances, evers2022theory,foo2025mind}. 
A leading mechanism is spin–orbit coupling (SOC) induced by molecular geometry~\cite{Savi2025chirality,chiesa_many-body_2024,PhysRevB.85.081404,PhysRevLett.108.218102,geyer2020effective}, which gives rise to interference between multiple transport pathways for electron transfer~\cite{guo_spin-dependent_2014,matityahu_spin-dependent_2016}. Building on this picture, many-body effects~\cite{Fransson2019,chiesa_many-body_2024} and environmental degrees of freedom~\cite{Das2022temperature,PhysRevB.102.235416,Ludena2025toward,Rudge2025role} have been shown to further modify the resulting spin-selective transport. 
A range of experimental studies have explored these effects in molecular systems and engineered structures, providing valuable insights into spin-dependent transport~\cite{calavalle2022gate,qian2022chiral,nakajima2023giant}. 
Synthetic quantum simulators, on the other hand, offer a highly controllable and versatile platform, where model Hamiltonians can be engineered and key parameters independently tuned, enabling systematic exploration of spin-dependent transport mechanisms.
While chirality-related phenomena have been explored in engineered quantum systems~\cite{roushan2017chiral,wang2019synthesis,deng2022observing,li2023observation,bouhiron2024realization,wang2024realizing,liang2024chiral,grass2025colloquium}, a direct simulation of electron transfer through chiral molecular bridges—capturing how the bridge structure governs quantum interference and gives rise to spin-dependent transport—remains largely unexplored. Trapped-ion systems provide a natural and programmable platform for this purpose~\cite{macdonell_analog_2021,kang_seeking_2024}, where multiple electronic sites are encoded in internal states~\cite{ringbauer2022universal, hrmo2023native} with precise control over coupling strengths and phases.
In addition, collective vibrational modes act as bosonic degrees of freedom~\cite{chen2021quantum}, which could introduce interactions affecting the electronic dynamics~\cite{PhysRevX.8.011038,so_trapped-ion_2024, so_quantum_2025}. These capabilities have enabled proof-of-principle simulations of molecular processes and provide a promising route toward controlled studies of chiral molecular transport.

Here we demonstrate a trapped-ion quantum simulation of spin-dependent electron transfer through a 
chiral molecular bridge. 
We consider a donor–bridge–acceptor architecture~\cite{chiesa_many-body_2024}, a model for electron transfer in which an excitation propagates from an initial donor state to a final acceptor state through intermediate bridge sites. We construct such an architecture with multiple energy-levels of the ion coupled by tunable laser and radio-frequency drives. In particular, the bridge consists of four sites with nearest-neighbor hopping and spin-dependent next-nearest-neighbor couplings that mimic SOC in molecules with chiral structure. We first investigate spin-dependent interference dynamics within the bridge and then study electron transfer from the donor state to the acceptor state mediated by the chiral structure. Finally, by tuning the coupling amplitude and phase within the bridge, we demonstrate controllable spin polarization in the transfer process. Our results establish a programmable platform for exploring spin-dependent transport in molecular systems with engineered internal structures.

We consider a 
chiral bridge model with four sites $|\mathrm{B}_1\rangle\sim|\mathrm{B}_4\rangle$ with nearest-neighbor and next-nearest-neighbor hopping. The donor and acceptor states, denoted as $|\mathrm{D}\rangle$ and $|\mathrm{A}\rangle$ respectively, are represented by two additional sites coupled to the ends of the bridge $|\mathrm{B}_1\rangle$ and $|\mathrm{B}_4\rangle$ respectively. 
These couplings are mediated by a common bath, modeled here as a single bosonic mode. The resulting donor–chiral bridge–acceptor (D–B$\chi$–A) Hamiltonian, illustrated schematically in Fig.~\ref{scheme}, is given by
\begin{equation}
\begin{aligned}
&H=H_\mathrm{B}+H_\mathrm{DB}+H_\mathrm{BA}, \\
    &H_\mathrm{B}=\sum_j\frac{t_je^{i\phi}}{2}|\mathrm{B}_{j+1}\rangle\langle \mathrm{B}_j|+\frac{i\lambda \sigma_z}{2}|\mathrm{B}_{j+2}\rangle\langle \mathrm{B}_j|+h.c.,\\
    &H_\mathrm{DB}=\frac{g_\mathrm{DB}}{2}(|\mathrm{D}\rangle\langle \mathrm{B}_1| +|\mathrm{B}_1\rangle\langle \mathrm{D}|) (a+a^\dagger),\\
    &H_\mathrm{BA}=\frac{g_\mathrm{BA}}{2}(|\mathrm{A}\rangle\langle \mathrm{B}_4| +|\mathrm{B}_4\rangle\langle \mathrm{A}|) (a+a^\dagger).
\end{aligned}
\end{equation}
Here, $t_j$ denotes the nearest-neighbor hopping strength, and $\phi$ is the phase that mimics the effect of an external magnetic field. Chirality is introduced via the next-nearest-neighbor SOC hopping with strength $\lambda \ge 0$ and the Pauli operator $\sigma_z$. 
For simplicity, We keep the spin polarization constant and simulate the transport of a spin-up (-down) electron with $\sigma_z=+1(-1)$ respectively. 
The parameters $g_\mathrm{DB}$ and $g_\mathrm{BA}$ describe the coupling strengths between the electron and the bosonic mode during donor–bridge and bridge–acceptor transfer, respectively, with $a$ the annihilation operator of the bosonic mode. As shown below, the transfer dynamics is governed by the chiral bridge structure, where quantum interference between different transport pathways leads to distinct
spin-dependent transfer probability at the acceptor, controlled by both the relative weight ratio $\lambda/t_j$ and the phase $\phi$.

\begin{figure}
  \centering
  \includegraphics[width=0.5\textwidth]{		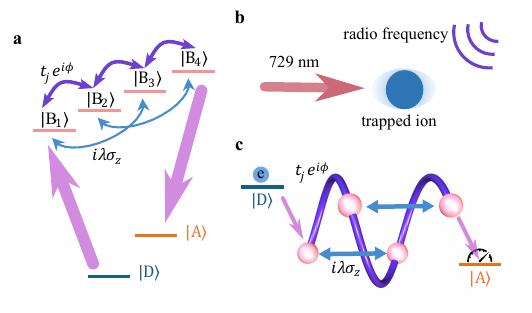}
  \caption{\textbf{Experimental scheme for simulating a donor–chiral bridge–acceptor system with a trapped ion.} \textbf{a}, Model of a donor–chiral bridge–acceptor system. 
  The bridge consists of four intermediate states $|\mathrm{B}_1\rangle\sim|\mathrm{B}_4\rangle$ with nearest-neighbor hopping $t_je^{i\phi}$ and spin-dependent next-nearest-neighbor hopping $i\mathbf{\lambda}\mathbf{\sigma_z}$ induced by SOC, with phase $\phi$ representing the effect of an external magnetic field. The donor $|\mathrm{D}\rangle$ and acceptor $|\mathrm{A}\rangle$ are coupled to the bridge via a bosonic mode. 
  \textbf{b}, Implementation of the couplings by 729~nm laser and radio-frequency drives, as detailed in the text. The laser also couples to the vibrational mode serving as the bosonic mode.
  \textbf{c}, Physical picture of chiral transport through the bridge. 
  The spin-dependent hopping generates interference pathways along the bridge, leading to spin-dependent transfer dynamics from the donor to the acceptor. The system is initialized in the donor state, and the population transferred to the acceptor is measured with varied coupling parameters.}  \label{scheme}
\end{figure}

Experimentally, we implement this model using a single trapped $^{40}\mathrm{Ca}^+$ ion~\cite{roos2000controlling}, as depicted in Fig.~\ref{scheme}b. Six internal levels within the $S_{1/2}$ and $D_{5/2}$ manifolds are used to encode the electronic sites: the two $S_{1/2,\pm1/2}$ states represent the donor and acceptor, while four $D_{5/2}$ levels ($m_j=\pm1/2, \pm3/2$) encode the bridge sites. The axial vibrational mode serves as the bosonic mode and is prepared close to the ground state via Doppler and sideband cooling, with a residual average phonon population of below 0.1. The donor–bridge and bridge–acceptor couplings are realized using a 729~nm laser aligned along the axial direction, which couples the internal states to the vibrational mode. Nearest-neighbor hopping within the bridge is implemented using a radio-frequency drive resonant with the energy splittings between the $D_{5/2}$ levels. Fixing $t_2=t$, the Clebsch–Gordan coefficients result in $t_1=t_3=\sqrt{8/9}t\approx t$. This specific realization yields dynamics close to the uniform-coupling case. 
To suppress unwanted coupling to the $D_{5/2,\pm5/2}$ states, additional off-resonant 729~nm beams are applied to induce AC Stark shifts, thereby shifting these levels out of resonance~\cite{li2025programmable}. The next-nearest-neighbor hopping is implemented via Raman transitions using two detuned 729~nm lasers~\cite{li2025beating}. All couplings are applied simultaneously, realizing an analog quantum simulation of the target Hamiltonian.
Population readout is performed using a sequential detection scheme~\cite{ringbauer2022universal}, in which each encoded state is mapped to the $S_{1/2}$ manifold and detected via fluorescence in sequence. See Supplemental Materials for further experimental details.

We first probe the intrinsic dynamics within the chiral bridge to isolate the interference mechanism. Starting from the left edge $|\mathrm{B}_1\rangle$, the population spreads across the bridge via competing nearest-neighbor and next-nearest-neighbor processes. The measured dynamics for $\lambda/t=1.42$ and $t=2\pi\times2.9$ kHz with $\phi=0$ are shown in Fig.~\ref{bridge}. At short times ($\sim50~\mu$s), the evolution is nearly spin independent: both spin components populate $|\mathrm{B}_2\rangle$ and $|\mathrm{B}_3\rangle$ with comparable amplitudes. Such phenomenon is attributed to the lowest-order process with single step nearest-neighbor hopping and next-nearest-neighbor steps without interference, thus not generating appreciable spin asymmetry. In contrast, at intermediate durations ($50\sim150~\mu$s), a pronounced spin separation appears with the spin-up component accumulating predominantly at $|\mathrm{B}_4\rangle$ (Fig.~\ref{bridge}b), while the spin-down component is reflecting back to $|\mathrm{B}_2\rangle$ (Fig.~\ref{bridge}c). As we show below, this behavior can be attributed to the interference between competing multi-step pathways. 
For even longer time scales, under coherent evolution, population dynamics exhibit oscillations.
We also show numerical simulations, with good agreement to experimental observations. 

To understand the interference dynamics, we consider the transition amplitude $A(\tau)=\langle \mathrm{B}_4|e^{-iH_\mathrm{B}\tau}|\mathrm{B}_1\rangle$. As an example for the interference dynamics, we consider a simply case with uniform hopping strength with $t_j=t$, with minor deviation from the implementation. We also focus on the case with $\phi=0$, with more general cases discussed below separately. Since $A(\tau)$ can be viewed as a coherent sum over all paths connecting the two end sites $|\mathrm{B}_1\rangle$ and $|\mathrm{B}_4\rangle$, each path involves a series of hopping steps with amplitude $\frac{t}{2}$ ($\frac{i\lambda\sigma_z}{2}$) for nearest-neighbor (next-nearest-neighbor) hopping. Starting from $|\mathrm{B}_1\rangle$, the lowest orders dynamics includes paths such as $\ket{\mathrm{B_1}}\to\ket{\mathrm{B_3}}\to\ket{\mathrm{B_4}}$,  $\ket{\mathrm{B_1}}\to\ket{\mathrm{B_2}}\to\ket{\mathrm{B_4}}$, $\ket{\mathrm{B_1}}\to\ket{\mathrm{B_2}}\to\ket{\mathrm{B_3}}\to\ket{\mathrm{B_4}}$ and $\ket{\mathrm{B_1}}\to\ket{\mathrm{B_3}}\to\ket{\mathrm{B_2}}\to\ket{\mathrm{B_4}}$. 
Building on these lowest-order processes, higher-order contributions can be understood as extensions of the same elementary paths, supplemented by additional segments that start and end on the same site and therefore factorized from the net propagation. As a result, the amplitude separates into two classes: even-order processes carry a factor $i\lambda\sigma_z$ from an odd number of next-nearest-neighbor hoppings, while odd-order processes contain a factor $i(\lambda^2 - t^2)$, reflecting the competition between direct and indirect tunneling pathways.
The interference between these two contributions gives rise to a term proportional to $\lambda\sigma_z(\lambda^2 - t^2)$, which governs the spin selectivity. In particular, this term vanishes at $\lambda=0$ and $\lambda=t$, corresponding to the absence of spin-dependent hopping and to destructive interference between competing pathways, respectively. Details are provided in Supplemental Materials.
\begin{figure}
  \centering
  \includegraphics[width=0.5\textwidth]{		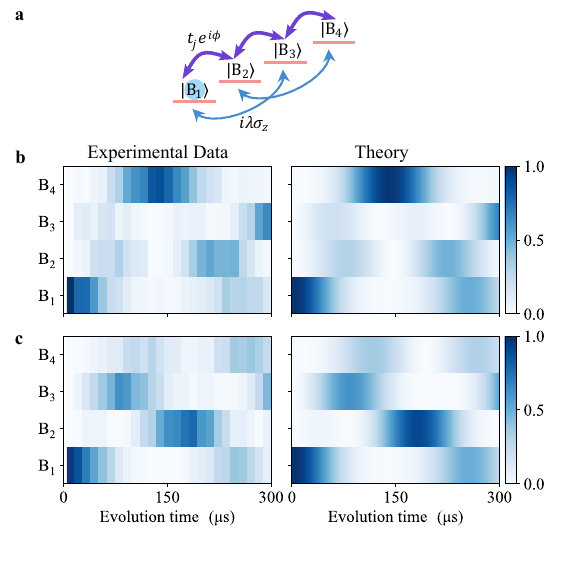}
  \caption{\textbf{Spin-dependent dynamics in the chiral bridge.} \textbf{a}, four-site chiral bridge $|\mathrm{B_1}\rangle\sim|\mathrm{B_4}\rangle$ with nearest-neighbor hopping $t_je^{i\phi}$ and spin-dependent next-nearest-neighbor hopping $i\lambda\sigma_z$, here we set $\phi=0$. 
  \textbf{b,c} Time evolution of site populations measured in the experiment (left) and obtained from numerical simulation considering the nonuniform couplings (right), starting from $|\mathrm{B_1}\rangle$. Each row corresponds to a bridge site. \textbf{b}, Spin-up dynamics. \textbf{c}, Spin-down dynamics. Parameters: $\lambda/t=1.42$, $t=2\pi\times2.9$ kHz.}  \label{bridge}
\end{figure}

Building on this understanding, we incorporate the donor and acceptor sites and investigate the full transport dynamics. As shown in Fig.~\ref{CISS}, the system is initialized in $|\mathrm{D}\rangle$ and evolves under the same bridge configuration. Because the donor–bridge and bridge–acceptor couplings ($g_\mathrm{DB}=g_\mathrm{BA}=2\pi\times2.8$ kHz) are comparable in magnitude to the intra-bridge hopping and are spin-independent, the overall dynamics remain governed by the bridge interference. The spin asymmetry observed within the bridge directly gives rise to the spin-selective transport within the time window $200$–$300~\mu\mathrm{s}$: the spin-up component, which preferentially reaches $|\mathrm{B}_4\rangle$, is efficiently transferred to the acceptor, while the spin-down component is partially reflected within the bridge and exhibits reduced transfer efficiency. This establishes a direct dynamical realization of chiral-induced spin selectivity. Amplitude analysis similar to the above case with only the bridge can be performed, as detailed in the Supplemental Materials.

\begin{figure}
  \centering
  \includegraphics[width=0.5\textwidth]{		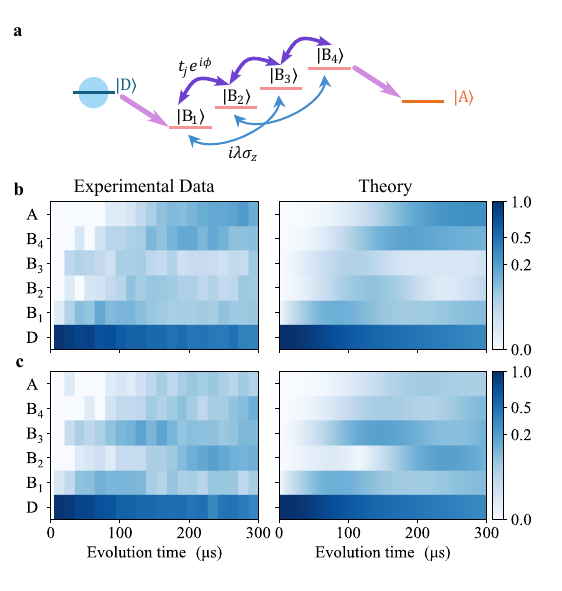}
  \caption{\textbf{Spin-dependent donor–acceptor transport mediated by the chiral bridge}. \textbf{a}, Donor–bridge–acceptor model. An electron initially in $|\mathrm{D}\rangle$ propagates through the four-site bridge $|\mathrm{B}_1\rangle\sim|\mathrm{B}_4\rangle$ and transfers to the acceptor state $|\mathrm{A}\rangle$. \textbf{b,c} Time evolution of site populations from experiment (left) and numerical simulation considering the realistic nonuniform couplings (right), with a distinction on the population at the acceptor state ($|\mathrm{A}\rangle$) over time. Each row represents one site. \textbf{b}, Spin-up dynamics. \textbf{c}, Spin-down dynamics. Parameters: 
$\phi=0$, $\lambda/t=1.4$, $t=2\pi\times2.9$ kHz, $g_\mathrm{DB}=g_\mathrm{BA}=2\pi\times2.8$~kHz.}  \label{CISS}
\end{figure}

Finally, we characterize the controllability of the spin polarization by tuning the bridge parameters. We focus on the dynamics around the first-arrival timescale and define the polarization as $P = (p_\uparrow - p_\downarrow)/(p_\uparrow + p_\downarrow)$, as a metric for the emergence of output polarization at the acceptor given a mixed donor input. 
We first vary the ratio $\lambda/t$, as shown in Fig.~\ref{Tunable}a and measure the dynamics at $253~\mu\mathrm{s}$. Theoretical results over the time window $200$–$300~\mu\mathrm{s}$ show a consistent polarization behavior, forming the shaded band, indicating that polarization is largely insensitive to time within this interval. A sign reversal of $P$ is observed around $\lambda/t\simeq1$, which is in agreement with the theoretical prediction of the interference term of the bridge $\propto \lambda\sigma_z(\lambda^2 - t^2)$. A small deviation from the ideal transition point arises from residual nonuniformity in the nearest-neighbor couplings, which is included in the numerical modeling and is found not to affect the underlying mechanism. This transition reflects the change from nearest-neighbor tunneling ($\lambda<t$) to next-nearest-neighbor-mediated loops ($\lambda>t$), and corresponds to the destructive interference point. 
We note that the experimentally accessible parameter range also extends to regimes relevant for typical chiral molecular systems, where $\lambda/t \ll 1$~\cite{chiesa_many-body_2024}, in which a finite spin polarization is still observed.

We further introduce a finite phase $\phi$ in the nearest-neighbor hopping (Fig.~\ref{Tunable}b), mimicking the gauge phase induced by an external magnetic field. We set $\lambda/t=1.4$ with other parameters kept the same as the above implementations except for varying $\phi$. The theory band, obtained over $200$–$300~\mu\mathrm{s}$, shows consistent behavior, indicating weak time dependence near the first-arrival timescale. The polarization exhibits a $\pi$-periodic modulation and vanishes near $\phi=\pi/4$. 
This can be understood by a gauge transformation $|\mathrm{B}_j\rangle \to e^{ij\pi/4} |\mathrm{B}_j\rangle$, which renders the Hamiltonian real and enforces time-reversal symmetry, thus removing spin selectivity. Details of the analysis is presented in the Supplemental Materials. The experimental data closely follow the theoretical curves, demonstrating controllability of spin selectivity via both coupling strengths and synthetic gauge phase.

\begin{figure}
  \centering
  \includegraphics[width=0.5\textwidth]{		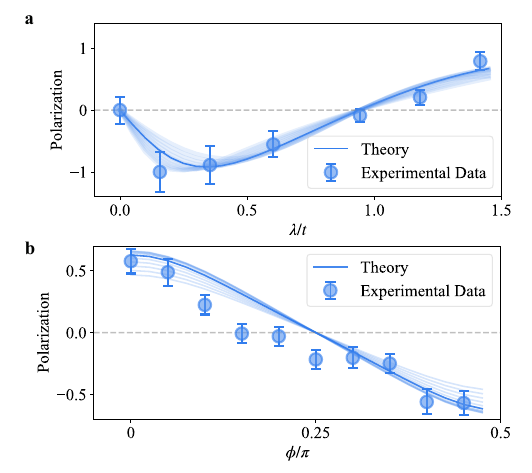}
  \caption{\textbf{Tunable spin polarization induced by the chiral bridge}.
\textbf{a}, Spin polarization as a function of the coupling ratio $\lambda/t$, measured at $253~\mu\mathrm{s}$. The polarization is negative for $\lambda/t<1$ and reverses sign for $\lambda/t>1$ ($\phi=0$). Theoretical curves are obtained over the time window $200$–$300~\mu\mathrm{s}$, showing consistent behavior within this interval.
\textbf{b}, Spin polarization as a function of the magnetic-field–induced hopping phase $\phi$ at $\lambda/t=1.4$, measured at $250~\mu\mathrm{s}$. The polarization changes sign with $\phi$ and approaches zero near $\phi=\pi/4$. Theoretical curves are obtained over $200$–$300~\mu\mathrm{s}$. Dots (solid blue lines) represent the experiment data (theory). The dashed grey line indicates the boundary between positive (above) and negative (below) spin polarization.}  \label{Tunable} 
\end{figure}

In conclusion, we demonstrate a programmable quantum simulation of spin-dependent electron transfer through a chiral bridge. 
We show that the quantum interference due to spin-dependent complex hopping and the multi-path structure of the bridge generates a spin-dependent transport, tunable by parameters including hopping amplitude and phase. 
Our results establish a framework in which chirality influences electron transfer through coherent pathway interference, providing an experimentally accessible mechanism for spin selectivity. Interestingly, our results show that spin-dependent complex phases, although rooted in chiral structure, do not by themselves produce spin selectivity; it is their interference across competing pathways that generates a measurable polarization. 
Our demonstration provides a route for exploring spin-dependent transport in complex chiral structure and environmental coupling. In particular, scaling to multi-ion systems enables access to multi-electron regimes for correlation effects in spin-selective transport~\cite{Fallas2025}. Incorporating multiple vibrational modes with engineered dissipation provides a pathway to study open-system dynamics and environment-assisted processes~\cite{so_trapped-ion_2024,sun2025quantum,so_quantum_2025}. Furthermore, programmable coupling geometries~\cite{Manovitz2020,Shapira2023} enable the realization of more complex bridge structures for interference-based mechanisms. These works may contribute to bridging programmable microscopic mechanisms and realistic molecular systems, toward a more thorough understanding of chiral-induced spin selectivity. 


We thank helpful discussion with Kenneth R. Brown, Qing-Feng Sun and Ting Rei Tan. The USTC team acknowledges support from the Chinese Academy of Sciences (Grant No.~XDB1300000), the National Natural Science Foundation of China (Grants No.~92565306, No.~22203083, No.~22373091), Quantum Science and Technology-National Science and Technology Major Project (Grant No.~2021ZD0301603, Grant No.~2021ZD0303303), the Fundamental Research Funds for the Central Universities (WK9990000169) and National Key Research and Development Program of China (Grant No.~2025YFE0217900). HP acknowledges support from the Welch Foundation (Grant No. C-1669). 


\clearpage{}

\onecolumngrid   
\newpage

\begin{center}
    \textbf{Supplemental Materials for:\\``Quantum Simulation of Spin-Dependent Electron Transfer in a Synthetic Chiral
Lattice with a Trapped Ion”}
\end{center}

\twocolumngrid   

\appendix
\setcounter{figure}{0}
\renewcommand{\thefigure}{S\arabic{figure}}  
\setcounter{table}{0}
\renewcommand{\thetable}{S\arabic{table}}    
\setcounter{equation}{0}
\renewcommand{\theequation}{S\arabic{equation}} 

\renewcommand{\theHfigure}{S.\arabic{figure}} 
\renewcommand{\theHtable}{S.\arabic{table}}
\renewcommand{\theHequation}{S.\arabic{equation}}


\maketitle

\section{Experimental implementation of the target Hamiltonian}

\begin{figure}
  \centering
  \includegraphics[width=0.5\textwidth]{		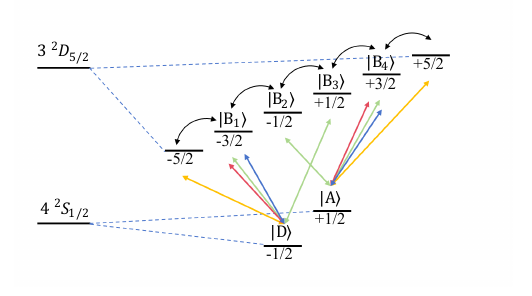}
  \caption{\textbf{Energy levels of $^{40}\mathrm{Ca}^+$ with control fields.} Red and blue arrows denote first-order sideband transitions generating spin-dependent forces. Green arrows indicate Raman couplings, black arrows represent radio-frequency driving, and orange arrows denote off-resonant beams used to shift unwanted levels.}\label{S1}
\end{figure}

We implement the model using a single trapped $^{40}\mathrm{Ca}^+$ ion in a linear Paul trap~\cite{RevModPhys.75.281}. 
The Hamiltonian of this model contains the donor–bridge and bridge–acceptor couplings, the nearest-neighbor hopping and the next-nearest-neighbor hopping. The donor–bridge and bridge–acceptor couplings are realized via spin-dependent forces generated by simultaneously driving red and blue sidebands (red and blue arrows in Fig.~\ref{S1}). This involves the bosonic axial motional mode of the ion, with frequency $\omega_m = 2\pi\times1.1$ MHz and Lamb–Dicke parameter $\eta = 0.0925$. The motional heating rate is about $0.1$ phonon/ms, which is negligible within the experimental duration (300 $\mu$s). The system operates well within the Lamb–Dicke regime with an average phonon number below 4.

Nearest-neighbor hopping within the bridge is implemented using an radio-frequency electric field that drives magnetic dipole transitions between adjacent Zeeman sub-levels in the $D_{5/2}$ manifold (black arrows in Fig.~\ref{S1}). Due to Clebsch–Gordan coefficients, the corresponding hopping strengths between $m$ and $m+1$ with $m = -5/2, -3/2, -1/2, +1/2, +3/2$ follow the ratio $\sqrt{5}:\sqrt{8}:\sqrt{9}:\sqrt{8}:\sqrt{5}$.
To suppress population leakage into the outermost $|D_{5/2,\pm5/2}\rangle$ states, we apply two laser beams off-resonant that detuned from the corresponding $|S_{1/2,\pm1/2}\rangle \leftrightarrow |D_{5/2,\pm5/2}\rangle$ transition to induce AC Stark shifts on these levels (orange arrows in Fig.~\ref{S1}). With Rabi frequency $\Omega_s=2\pi\times0.1$ MHz and detuning $\delta_s=2\pi\times0.5$ MHz for both laser beams, the AC Stark shift $\Omega_s^2/(4\delta_s)\approx2\pi\times0.005$ MHz. Both numerical simulations and experiments show that the residual populations in $|D_{5/2,+5/2}\rangle$ and $|D_{5/2,-5/2}\rangle$ during the dynamics remain below 3\% and 5\%, respectively. Therefore, this residual population leakage does not affect the experimental observations. For simplicity, we neglect the population in $|D_{5/2,+5/2}\rangle$ for population readout.

Next-nearest-neighbor hopping within the chiral bridge is engineered using two pairs of off-resonant 729~nm Raman beams (green arrows in Fig.~\ref{S1}). In particular, the coupling $|D_{5/2,-3/2}\rangle \leftrightarrow |D_{5/2,+1/2}\rangle$ is mediated via $|S_{1/2,-1/2}\rangle$, while $|D_{5/2,-1/2}\rangle \leftrightarrow |D_{5/2,+3/2}\rangle$ is mediated via $|S_{1/2,+1/2}\rangle$. All Raman beams share the same Rabi frequency $\Omega_2=2\pi\times0.041$ MHz and detuning $\delta$, resulting in an effective hopping strength $\Omega_2^2/(2\delta)$.
In the main text (Fig.~2, Fig.~3, and Fig.~4(b)), the detuning $\delta$ is set to $2\pi\times0.2$ MHz. In Fig.~4(a), we vary the next-nearest-neighbor coupling strength $\lambda$ by scanning the detuning $\delta$, except at the $\lambda=0$ point where all Raman lasers are turned off. 

\vspace{4\baselineskip} 

\section{State preparation and readout}
\textit{State preparation.—}
The ion's internal state is initialized to the donor state $|S_{1/2,-1/2}\rangle$ using frequency-resolved optical pumping with the 729 nm laser, assisted by 854 nm and 866 nm repump lasers. The bosonic mode is prepared close to the ground state via Doppler and sideband cooling, with a residual phonon occupation below 0.1.

\textit{Population readout.—}
We employ a sequential fluorescence detection scheme to extract the populations of multiple sub-levels~\cite{ringbauer2022universal}.
The detection sequence proceeds as follows. First, a shelving $\pi$ pulse is applied to transfer population  $|S_{1/2,+1/2}\rangle\rightarrow|D_{5/2,+5/2}\rangle$, leaving only $|S_{1/2,-1/2}\rangle$ (donor state $ |\mathrm{D}\rangle$ in the text) occupied in the $S_{1/2}$ manifold. Then, we use 397 nm and 866 nm lasers to perform a standard fluorescence readout for the states in the ground manifold, allowing us to identify the population in $|\mathrm{D}\rangle$ state without perturbing the rest. 
Next, the same $\pi$ pulse is applied again to transfer the original population of $|S_{1/2,+1/2}\rangle$ ($|\mathrm{A}\rangle$ state)  back to the $S_{1/2}$ manifold. Then, we perform the fluorescence readout to identify the $|\mathrm{A}\rangle$ state.
Subsequently, we sequentially apply $\pi$ pulses that map $|\mathrm{B_1}\rangle$, $|\mathrm{B_2}\rangle$, $|\mathrm{B_3}\rangle$, and $|\mathrm{B_4}\rangle$ onto the $S_{1/2}$ manifold, performing a fluorescence readout after each mapping. In this way, all six sub-levels are read out within a single experimental run. In this sequence, the first occurrence of a bright fluorescence signal during the repeated readouts indicates that the system has collapsed onto the corresponding state, and averaging over many experimental runs yields the population of each state.

Within the experimental timescale of 300 $\mu$s, all $\pi$ pulses used in the detection sequence achieve fidelities above 97\%. For longer evolution times, motional heating degrades the fidelity, thereby limiting the accessible observation window. This window nevertheless allows us to study the first-arrival-stage dynamics, which captures the physics of electron transfer. 

\section{Theoretical analysis of the multi-path interference}


\begin{figure}[ht]
  \centering
  \includegraphics[width=0.5\textwidth]{		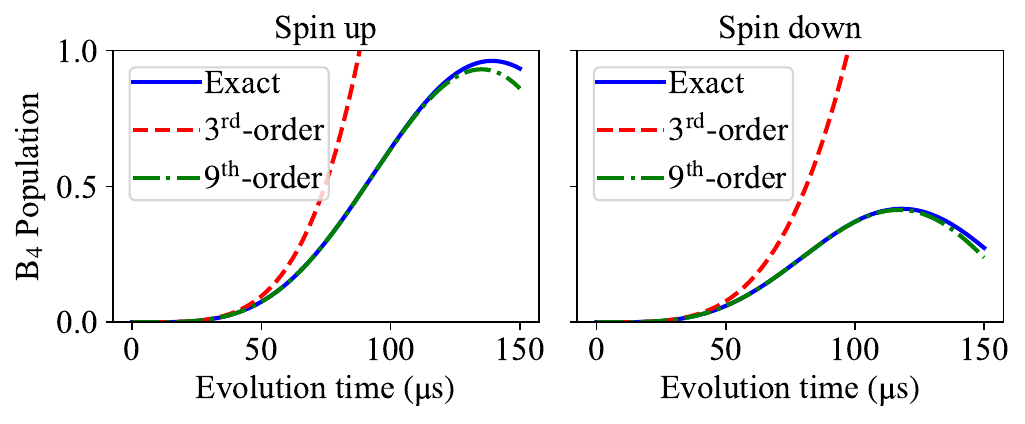}
  \caption{\textbf{Time evolution of B$_4$ state population.} Exact diagonalization results are shown as a blue solid curve. Taylor expansions up to third order (red dashed) and ninth order (green dash-dotted) are compared. The nighth-order expansion accurately reproduces the exact dynamics for time $0 \sim 150$ $\mu$s, confirming the validity of the perturbative approach in the first-arrival timescale.}\label{S2}
\end{figure}


In this Appendix, we provide a path-resolved decomposition of the transition amplitude
$A_{1\to4}(\tau)=\langle \mathrm{B}_4|e^{-iH_\mathrm{B}\tau}|\mathrm{B}_1\rangle$, 
which underlies the interference mechanism discussed in the main text, with the bridge Hamiltonian following Eq.~1. Rather than relying on a low-order truncation, we organize the expansion according to the parity of hopping processes, which captures the relevant structure of the dynamics over the experimentally probed timescales. Here we consider $\phi=0$ for simplicity.

We expand the evolution operator in powers of $\tau$, each term corresponds to a sum over paths with a given number of hopping steps. Owing to the structure of the Hamiltonian, contributions naturally separate into two classes for odd and even number of next-nearest-neighbor hoppings, as described in the text. 
The lowest order terms are listed in Tab.~\ref{tab:path}. Higher order processes are the combination of second and third order process.
This allows the amplitude to be written in the form
$A_{1\to4}(\tau)= i\lambda\sigma_z f(\tau,\lambda,t)+i(\lambda^2 - t^2)g(\tau,\lambda,t)$
where $f$ and $g$ are real-valued functions collecting contributions from the corresponding path families. Within the parameter regime and timescales considered here, both functions remain positive (when expand $A_{1\to4}$ to ninth order in $\tau$), so that the relative sign between the two terms is determined solely by $(\lambda^2 - t^2)$, which in turn sets the sign of the spin polarization.
We compare this decomposition with exact diagonalization in Fig.~\ref{S2}. We find that agreement over the relevant timescale requires inclusion of higher-order processes (up to ninth order in $\tau$), while the above parity-based structure remains valid and accurately captures the observed spin-dependent dynamics.

For a finite phase $\phi$, the hopping amplitudes acquire additional complex phases, making a full path-resolved expansion at higher orders impractical. Instead, the phase dependence can be understood from symmetry considerations of the Hamiltonian.
Applying a gauge transformation $|\mathrm{B}_j\rangle \to e^{ij\phi} |\mathrm{B}_j\rangle$, the Hamiltonian becomes
\begin{equation}
H_\mathrm{B}'=\sum_j\frac{t_j}{2}|\mathrm{B}_{j+1}\rangle\langle \mathrm{B}_j|+\frac{i\lambda \sigma_z e^{-i2\phi}}{2}|\mathrm{B}_{j+2}\rangle\langle \mathrm{B}_j|+\mathrm{h.c.},    
\end{equation}
from which it follows that $H_\mathrm{B}'$ is $\pi$-periodic in $\phi$. Consequently, the spin polarization also exhibits $\pi$-periodicity.

At the special point $\phi=\pi/4$, the remaining complex phase can be factored out such that the Hamiltonian becomes real-valued. In this case, a staggered transformation $|\mathrm{B}_j\rangle \to (-1)^j |\mathrm{B}_j\rangle$ maps $H_\mathrm{B}'(\sigma_z)$ to $-H_\mathrm{B}'(-\sigma_z)$. Combining these two properties, the transition amplitudes satisfy
\begin{equation}
\begin{aligned}
    A_{1\to4}(\sigma_z,\tau)&=\langle \mathrm{B}_4|e^{-iH_\mathrm{B}'(\sigma_z)\tau}|\mathrm{B}_1\rangle\\
    &=\langle \mathrm{B}_4|e^{iH_\mathrm{B}'(-\sigma_z)\tau}|\mathrm{B}_1\rangle\\
    &=\langle \mathrm{B}_4|e^{-iH_\mathrm{B}'(-\sigma_z)\tau}|\mathrm{B}_1\rangle^*\\
    &=A_{1\to4}^*(-\sigma_z,\tau),
\end{aligned}
\end{equation}
where in the third line we used that the Hamiltonian is real. Therefore, the transition amplitudes for opposite spins are related by complex conjugation. As a result, $|A_{1\to4}(\sigma_z,\tau)|^2=|A_{1\to4}(-\sigma_z,\tau)|^2$,
the population transport to $\mathrm{B}_4$ becomes independent of $\sigma_z$, and the spin polarization vanishes. This explains the disappearance of spin selectivity at $\phi=\pi/4$ observed in the main text.

\section{Interference analysis with motional mode}
Consider an initial state  $\rho(\tau=0)=\ket{\mathrm{D}}\bra{\mathrm{D}}\rho_m$, with $\rho_m$ the initial motional density matrix, the population in the acceptor is then $P=Tr[\ket{A}\bra{A}e^{-iH\tau}\ket{\mathrm{D}}\bra{\mathrm{D}}\rho_m e^{iH\tau}]$. One can introduce the position eigen-basis $\{\ket{x}\}$ for the ion motion, then the Hamiltonian can be parametrized by the position value as $H=\int_{-\infty}^{+\infty} H(x)\ket{x}\bra{x}d x$, since $a+a^\dagger\propto \int_{-\infty}^{+\infty} x\ket{x}\bra{x}d x$. With the diagonal components of the initial motional density matrix $diag(\rho_m)=\int_{-\infty}^{+\infty} \rho(x)\ket{x}\bra{x}d x$, thus $P=\int_{-\infty}^{+\infty} |\bra{\mathrm{A}}e^{-i H(x)\tau}\ket{\mathrm{D}}|^2 \rho(x)d x $, as a weighted integral of the evolution over the discrete space within $H(x)$ from donor to acceptor. Then the amplitude components from the donor to the acceptor can be expanded in a series of $\tau$ as a result of the interferences of possible routes, similar to the above analysis with only the bridge.

\renewcommand{\arraystretch}{1.8}
\begin{table}[ht]
\centering
\caption{\textbf{Path decomposition of the transition amplitude up to the third order.} Each path corresponds to a sequence of hoppings between intermediate sites, with amplitudes determined by the hopping matrix elements and the coefficients from the Taylor expansion. The lower part of the table lists the resulting terms in $p_\sigma = |A_2 + A_3|^2$, highlighting the interference contribution responsible for the spin dependence.}\label{tab:path}
\begin{tabular}{|l|c|c|}
\hline
\textbf{Transfer path} & \textbf{Order} & \textbf{Amplitude} \\
\hline
$\mathrm{B_1\to B_3\to B_4}$ & 2nd & ${-ite^{i\phi} \lambda\sigma_z \tau^2}/{8}$ \\
\hline
$\mathrm{B_1\to B_2\to B_4}$ & 2nd & ${-ite^{i\phi} \lambda\sigma_z \tau^2}/{8}$ \\
\hline
\textbf{Total 2nd-order} & 2nd & $A_2 = {-it e^{i\phi} \lambda\sigma_z \tau^2}/{4}$ \\
\hline
 $\mathrm{B_1\to B_2\to B_3\to B_4}$ & 3rd & $it^3 e^{3i\phi}\tau^3/48$ \\
\hline
$\mathrm{B_1\to B_3\to B_2\to B_4}$ & 3rd & $-it e^{-i\phi}\lambda^2\tau^3/48$ \\
\hline
\textbf{Total 3rd-order} & 3rd & $A_3 = it e^{-i\phi}(t^2 e^{4i\phi} - \lambda^2)\tau^3/48$ \\
\hline
\multicolumn{3}{|l|}{\textbf{Non-interference terms in probability:}} \\
\hline
\multicolumn{2}{|c|}{$|A_2|^2$} & $t^2\lambda^2\tau^4/16$ \\
\hline
\multicolumn{2}{|c|}{$|A_3|^2$} & $(t^4+\lambda^4-2t^2\lambda^2\cos 4\phi)t^2\tau^6/2304$\\
\hline
\multicolumn{3}{|l|}{\textbf{Spin-selective interference term (up to the 3rd-order):}} \\
\hline
\multicolumn{2}{|c|}{$2\mathrm{Re}(A_2 A_3^*)$} & $t^2\lambda(\lambda^2-t^2)\sigma_z\cos(2\phi){\tau^5}/96$ \\
\hline
\end{tabular}
\end{table}
%

\clearpage{}

\end{document}